\documentclass[11pt]{article}

\usepackage{amsmath,amsfonts,amssymb,amsthm}
\usepackage[margin=1in]{geometry}
\usepackage[colorlinks]{hyperref}
\usepackage{graphicx}

\begin{document}


\title{A Revisit to Non-maximally Entangled Mixed States: Teleportation Witness, Noisy Channel and Discord}

\author{Sovik Roy$^{1,2}$ \\
$^1$ Department of Mathematics, EM 4/1, Techno India, Salt Lake, Kolkata - 91, India\\
$^2$ S. N. Bose National Centre for Basic Sciences, Salt Lake, Kolkata - 98, India\\\\
Biplab Ghosh$^{3}$ \\
$^3$ Department of Physics, Vivekananda College for Women, Barisha, Kolkata - 8, India}

\date{\today}
\maketitle

\begin{abstract}
\noindent We constructed a class of non-maximally entangled mixed states \cite{roy2010} and extensively studied its entanglement properties and also their usefulness as teleportation channels. In this article, we revisited our constructed state and have studied it from three different perspectives. Since every entangled state is associated with an witness operator, we have found a suitable entanglement as well as teleportation witness for our non-maximally entangled mixed states. We considered the noisy channel's effects on our constructed state and to see whether it affects the states' capacity as teleportation channel. For this purpose we have mainly emphasized on amplitude damping channel. A comparative study with concurrence and quantum discord of the state of ref. \cite{roy2010} has also been carried out here.
\end{abstract}

\section{Introduction}
\label{sec:intro}

Quantum entanglement has always been a key ingredient ever since its inception and fundamentally motivates in fulfilling several interesting tasks in the domain of quantum information processing. Such tasks involve teleportation \cite{bennett1993}, dense coding \cite{bennett1992}, secret sharing \cite{hillery1999}, quantum cryptography \cite{bennett1984} and many others. Amongst all the aforementioned tasks we shall emphasize specifically on teleportation in this article. Ideally using pure maximally entangled states, like singlet state ($\frac{\vert 01\rangle + \vert 10\rangle}{\sqrt{2}}$), one would always expect to perform such a task with high degree of fidelity. But due to the de-coherence (or noise) effect of nature, one encounters with mixed entangled state mostly. For $2\otimes 2$ dimension, Werner state \cite{werner1989}, which is a maximally entangled mixed state, was shown to be a useful resource  in teleportation. The average optimal teleportation fidelity corresponding to this state exceeds classical teleportation fidelity $\frac{2}{3}$ \cite{lee2000}. Another important class of maximally entangled mixed states is by Munro \textit{et. al} \cite{munro2001}. We earlier showed that, Munro states could be used as a faithful teleportation channel only when the mixedness of the state is less than some specific values while for all finite degrees of mixedness, average optimal teleportation fidelity of Werner state exceeds that of Munro states \cite{roy2010}. Our central point of investigation here is non-maximally entangled mixed class of states. Those states which achieve the greatest possible entanglement for a given degree of mixedness are basically known as maximally entangled mixed states (MEMS) but the states which cannot be categorized in this way are called non-maximally entangled mixed states (NMEMS). One of the well known members of non-maximally entangled mixed state class is Werner derivative, first introduced by Ishizaka and Hiroshima \cite{hiroshima2000}. Werner derivative states were also shown to be useful as teleportation channels depending upon some of the state's parameter ranges \cite{roy2010}. We, in this paper, make a revisit to the domain of a special non-maximally entangled mixed state which we, along with several of other co-authors, constructed in one of our earlier works of \cite{roy2010}.
We studied the states' fidelity of teleportation also and their capacities as teleportation channels were compared with that of different MEMS and NMEMS (Werner derivative). We also found that our NMEMS, as teleportation channels, outperformed Werner derivative. The motivation for our present work is to look at the constructed NMEMS from different perspectives. We shall try to find a suitable teleportation witness operator for the constructed NMEMS as well as to see how such NMEMS behaves in noisy environment. Also quantum discord, which is a measurment of quantum correlation has also been considered here. We have made a comparative study of two measures (such as concurrence and quantum discord) by taking into consideration our constructed NMEMS class. To begin with, in section $2$ we briefly discuss about our NMEMS first and then about teleportation witness operators. Section $3$ discusses about the suitable teleportation witness for our NMEMS. In section $4$ however we have studied the effects of noisy channel on NMEMS and have analyzed the measurement induced disturbance imposed on the NMEMS by noisy channel like amplitude damping channel. We wanted to see how these channels affect the teleportation fidelities of the class of NMEMS. In section $5$ quantum discord of the constructed NMEMS has been studied and this measure has been compared with measurement of entanglement viz. concurrence. The paper concludes with section $6$.
\section{Pre-requisites: a brief recapitulation}
\label{sec:prereq}
\subsection*{Proposed NMEMS:}
The proposed two qubit NMEMS was defined as follows \cite{roy2010}.
\begin{eqnarray}
\rho_{nmems} = p\:\rho_{ab}^{G} + (1-p)\:\rho_{ab}^{W},\:\:\:\:
0\:\leq\:p\:\leq\:1.
\label{p_1}
\end{eqnarray} 
where, $\rho_{ab}^{G} = Tr_{c}(\vert GHZ\rangle_{abc})$ (a separable density matrix) and $\rho_{ab}^{W} = Tr_{c}(\vert W\rangle_{abc})$ (an inseparable density matrix). $\vert GHZ\rangle_{abc}$ and $\vert W\rangle_{abc}$ are known as $GHZ$ state and $W$ state respectively \cite{ghz1989}. The characteristics of $GHZ$ and $W$ states are that, they respectively are two qubit separable and inseparable states when a qubit is lost from the corresponding three qubit states. The state (\ref{p_1}) was shown to be entangled when $0\:\leq\: p\:< \:0.292$ but is useful in teleportation only when $0\:\leq\: p\:< \:0.25$. The state however could not be used in teleportation when $0.25\:\leq\: p\:< \:0.292$, although it remained entangled over there. It is a known fact that, though the quantum correlations violate Bell's inequality, they satisfy weaker inequalities of a similar type such as Cirelson's bound \cite{cirelson1980}. Poh \textit{et. al} have recently proved that, satisfying the Cirelson's bound can also be used to certify that the state under consideration is a maximally entangled state \cite{poh2015}. It has already been shown in \cite{roy2010} that the state $\rho_{nmems}$, for entire region of $p$ i.e. $ 0\:\leq p\: \leq 1$ does not violate Bell-CHSH inequality although it is entangled for $0\:\leq p \: <0.292$. Since Cirelson's bound is a weaker inequality in comparison to that of Bell-CHSH inequality, from the above analysis we can interpret that the state $\rho_{nmems}$ satisfies the Cirelson's bound too. Indeed, the state (\ref{p_1}) reduces to the form 
\begin{eqnarray}
\vert \varrho\rangle = \frac{1}{3}\:\vert 00\rangle_{ab}\langle 00\vert + \frac{2}{3}\:\vert \Psi^{+}\rangle_{ab}\langle \Psi^{+}\vert, 
\label{p_2}
\end{eqnarray}
which is maximally entangled as it can be put into Ishizaka - Hiroshima proposed class of MEMS \cite{ishizaka2000} and this happens when the parameter of the state (\ref{p_1}) i.e. $p$ takes the limiting value as zero. Here $\vert \Psi^{+}\rangle = \frac{\vert 01\rangle + \vert 10\rangle}{\sqrt{2}}$.
\subsection*{Teleportation witness operator:}
As we know that the pivotal ingredient for performing teleportation task is accentuated by quantum entanglement, the existence of entanglement is however authenticated by its detection by means of entanglement witness. An observable $W_{t}$ is called entanglement witness if (i) $Tr(W_{t}\varrho_{s})\geq 0$ for all separable $\varrho_{s}$ and (ii) $Tr(W_{t}\varrho_{e})< 0$ for atleast one entangled $\varrho_{e}$. It is also a known fact that for each entangled state $\varrho_{e}$ there exists a suitable entanglement witness for its detection \cite{guhne2009}. Ganguly \textit{et. al} investigated another aspect of this interesting facet of witness. They proved the existence of teleportation witnesses which signified whether an unknown entangled state (pure or mixed) could be used as a resource in teleportation \cite{ganguly2011}. A hermitian operator $W_{s}$ may be called a teleportation witness if the following conditions hold good, (a) $Tr(W_{s}\sigma)\geq 0$, for all states $\sigma$ which are not useful for teleportation, (b) $Tr(W_{s}\chi)< 0$ for atleast one state $\chi$ which is useful for teleportation. For $d\otimes d$ dimensional system the teleportation witness proposed by Ganguly \textit{et. al} is 
\begin{eqnarray}
W_{s} = \frac{1}{d}\:I - \vert \psi^{+}\rangle\langle \psi^{+}\vert
\label{p_3},
\end{eqnarray}
where $\vert \psi^{+}\rangle = \frac{1}{\sqrt{d}}\:\sum_{i=0}^{d-1}\vert ii\rangle$.
\section{Teleportation witness for NMEMS:}
The main purpose of this revisit to our proposed class of NMEMS is to look for a suitable teleportation witness which will further validate the observational inference attained about the state (\ref{p_1}) regarding its feasibility in teleportation. In computational basis the state (\ref{p_1}) is expressed as
\begin{eqnarray}
\rho_{nmems} = \left(%
\begin{array}{cccc}
\frac{p+2}{6} & 0 & 0 & 0 \\
0& \frac{1-p}{3} & \frac{1-p}{3} & 0\\
0 & \frac{1-p}{3}&\frac{1-p}{3} & 0 \\
 0 & 0 & 0 & \frac{p}{2}
\end{array}%
\right)
\label{p_4}.
\end{eqnarray}
The witness defined in (\ref{p_3}) cannot detect the applicability of the state (\ref{p_1}) in teleportation as it is found that in this case $Tr(W_{s}\:\rho_{nmems})=\frac{1-p}{3}\:>\: 0$, (here $d=2$). The state (\ref{p_1}) represents a class of non-maximally entangled mixed state and from eqs. (\ref{p_1}) and (\ref{p_4}) it is clear that the NMEMS is parametrized by the states' parameter $p$. It is also clear from ref. \cite{roy2010} that not for all $p$'s the states are useful as teleportation channels. Their utilities as teleportation channels rather depend on specific parameter ranges. The state (\ref{p_1}) is useful in teleportation when $0\:\leq\:p\:<\:0.25$. So here we begin to look for a suitable teleportation witness for our constructed state $\rho_{nmems}$. It has been shown in \cite{guhne2009} that, a witness for genuine multi-partite entanglement around the state $\vert W\rangle_{abc}$ would be of the form $\frac{2}{3}\:I-\vert W\rangle_{abc}\langle W\vert$, while a witness of the form $\frac{4}{9}\:I-\vert W\rangle_{abc}\langle W\vert$ would exclude full separability. Now if one considers a witness operator of the form
\begin{eqnarray}
W^{(1)}_{t} = \frac{4}{9}\:I - \rho_{ab}^{W}
\label{p_5},
\end{eqnarray}
where, $\rho_{ab}^{W}= Tr_{c}(\vert W\rangle_{ab}\langle W\vert)$, then it is observed that $Tr(W_{t}^{(1)}\:\rho_{nmems}) = \frac{7\:p - 2}{18}$ and hence the witness successfully detects the entanglement of the state $\rho_{nmems}$ for $0\:\leq p\: \leq 0.2857$. It is interesting to observe that the eq. (\ref{p_5}) represents a suitable entanglement witness for the state $\rho_{nmems}$ and this tallies with our earlier observation from ref. \cite{roy2010}. But the witness (\ref{p_5}) cannot be considered as a suitable teleportation witness for the state $\rho_{nmems}$ as we know that in the range $0.25 \: \leq p \: < 0.292$ the state although entangled cannot be used in teleportation. Amongst all witnesses the stabilizer witnesses are very convenient for doing experiments \cite{guhne2009}. We now construct one such witness as follows
\begin{eqnarray}
W^{(2)}_{t} = I - \sigma_{x}\otimes \sigma_{x} - \sigma_{y}\otimes \sigma_{y}
\label{p_6},
\end{eqnarray}
where, $\sigma_{x}$ and $\sigma_{y}$ are Pauli spin matrices. It is then observed that $Tr(W^{(2)}_{t}\:\rho_{nmems}) = \frac{4\:p-1}{3}$, which is negative only when $0\:\leq p\: < 0.25$. Thus, the witness (\ref{p_6}) is the more sought after as it is capable of detecting that, the state (\ref{p_1}) is useful in teleportation.
\section{Amplitude damping channel and NMEMS:}
The basic idea of ref. \cite{roy2010} was to prove the usefulness of the state $\rho_{nmems}$ as teleportation channel and how this depends on the choice of appropriate ranges of state parameter. The idea, in this section, is to study about how environmental noises affect such channels. When the environment-channel correlation time is comparatively smaller, such simplest models are termed as memoryless while the models are said to have memory when the correlation time between channel and environment is significant. Amplitude damping channels are one such environment-channel interaction models with memory. These models are used to characterize the description of energy dissipation-effects due to loss of spontaneous emission energy from a quantum system \cite{nielsenbook}. For an arbitrary single qubit $\rho$, if we denote the amplitude damping channel (ADC) by $\varepsilon_{AD}$, then the effects of ADC on $\rho$ is mathematically represented as
\begin{eqnarray}
\varepsilon_{AD}\:(\rho) = E_{0}\rho\:E_{0}^{\dagger} + E_{1}\rho\:E_{1}^{\dagger}
\label{p_7},
\end{eqnarray}
where,
\begin{eqnarray}
 E_{0}= \left(%
\begin{array}{cc}
1 & 0  \\
0& \sqrt{1-\gamma} 
\end{array}%
\right),\:\:\:\:
 E_{1}= \left(%
\begin{array}{cc}
0 & \sqrt{\gamma} \\
0& 0 
\end{array}%
\right)
\label{p_8},
\end{eqnarray}
where $E_{0}$ and $E_{1}$ are Kraus' operators \cite{Krausbook}. For two qubit states like $\rho_{nmems}$, the ADC channel effects are given as
\begin{eqnarray}
\varepsilon_{AD}\:(\rho_{nmems}) = (E_{0}\otimes E_{0})\rho_{nmems}\:(E_{0}\otimes E_{0})^{\dagger} + (E_{1}\otimes E_{1})\rho_{nmems}\:(E_{1}\otimes E_{1})^{\dagger}
\label{p_9}.
\end{eqnarray}
If we consider $\gamma = \sin^{2}(\theta)$, then using eq. (\ref{p_9}), the state (\ref{p_1}) transforms and   in computational basis the transformed state  is shown to be
 \begin{eqnarray}
\rho_{nmems}^{AD} = \left(%
\begin{array}{cccc}
\frac{p+2}{6} & 0 & 0 & 0 \\
0& \frac{1-p}{3}\:(1-\gamma) & \frac{1-p}{3}\:(1-\gamma) & 0\\
0 & \frac{1-p}{3}\:(1-\gamma)&\frac{1-p}{3}\:(1-\gamma) & 0 \\
 0 & 0 & 0 & \frac{p}{2}\:(1-\gamma)^{2}
\end{array}%
\right)
\label{p_10},
\end{eqnarray}
where $\rho_{nmems}^{AD}$ is the transformed $\rho_{nmems}$ after noise effects. Now for an arbitrary state $\varsigma$, whose density matrix is of the form
\begin{eqnarray}
\varsigma = \left(%
\begin{array}{cccc}
a & 0 & 0 & 0 \\
0& b & c & 0\\
0 & c^{*} &d & 0 \\
 0 & 0 & 0 & e
\end{array}%
\right)
\label{p_11},
\end{eqnarray}
the amount of entanglement (which is quantified by $`$concurrence') is given by \cite{bruss2003}
\begin{eqnarray}
C(\varsigma) = 2\:\max\:(\:\vert c\vert - \sqrt{a\:e},0\:)
\label{p_12}.
\end{eqnarray}
Using eq. (\ref{p_12}), the concurrence of the state $\rho_{nmems}$ was found to be \cite{roy2010}
\begin{eqnarray}
C(\rho_{nmems}) = 2\:\max\:(\:\frac{1-p}{3} - \sqrt{\frac{p\:(p+2)}{12}},0\:). 
\label{p_13}
\end{eqnarray}
Therefore, $\rho_{nmems}$ is entangled only if $\frac{1-p}{3} - \sqrt{\frac{p\:(p+2)}{12}}\:>\:0$, i.e. when $0\:\leq p\:<0.292$. It is also clear that the state (\ref{p_10}) also falls into the category of density matrices represented by (\ref{p_11}). So using eq. (\ref{p_12}), the concurrence of the state $\rho_{nmems}^{AD}$ is obtained as
\begin{eqnarray}
C(\rho_{nmems}^{AD}) = (1-\gamma)\:C(\rho_{nmems}),\:\:\:\:\gamma=\sin^{2}(\theta).
\label{p_14}
\end{eqnarray}
Below, we plot a figure to show how concurrences vary for the states (\ref{p_1}) and (\ref{p_10}).
\begin{figure}[h!]
\begin{center}
\includegraphics[width=11cm]{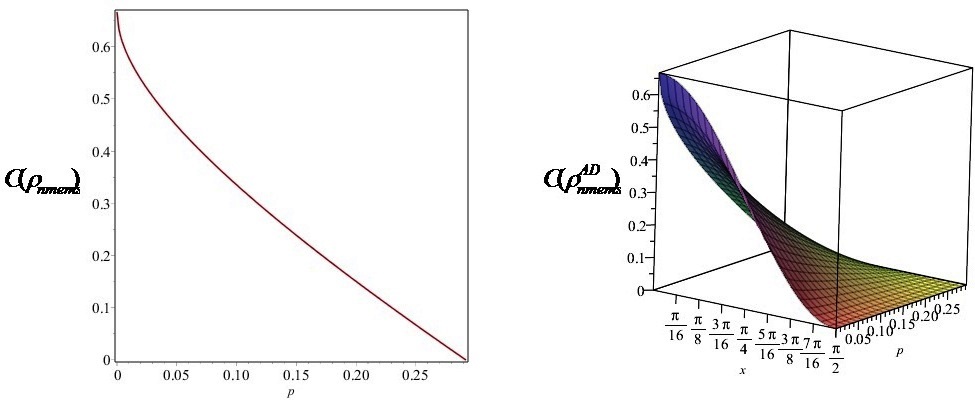}
\caption{(Coloronline) The first figure of the plot represents the nature of entanglement of the state $\rho_{nmems}$ while the second figure shows the entanglement of the state $\rho_{nmems}^{AD}$. In both these cases $0\:\leq p\:<\:0.292$.}
\end{center}
\label{f1}
\end{figure}
It is clear from Fig. $1$ that when $p=0$ and $\theta=0$, then $C(\rho_{nmems}^{AD})=C(\rho_{nmems})$, whereas when $p=0$ and $\theta=\frac{\pi}{2}$, the concurrence of state $\rho_{nmems}^{AD}$ vanishes. However for $p=0$ and $\theta=\frac{\pi}{4}$, the concurrence of the state $\rho_{nmems}^{AD}$ is $0.3333$ and so on. It is therefore obvious from the above analysis that the concurrence is a decreasing function of the state parameter $p$. Amplitude damping channel considerably affects the entanglement of the state (\ref{p_1}). Originally the state $\rho_{nmems}$ could be used as teleportation channels for specified range of state parameter $p$. The teleportation fidelity of the state (\ref{p_1}) was found to be 
\begin{eqnarray}
f^{T}_{opt}\:(\rho_{nmems}) = \frac{7-4\:p}{9},\:\:\:\: 0\:\leq p\:<\:0.25
\label{p_15}.
\end{eqnarray}
It was also seen that the state's teleportation fidelity is 0.77 (for $p=0$) which exceeds classical teleportation fidelity of $\frac{2}{3}$ \cite{roy2010}. Amplitude damping channel affects the entanglement of the state $\rho_{nmems}$ and the teleportation fidelity of the state reduces thereby. The teleportation fidelity of the state $\rho_{nmems}^{AD}$ is however given by
\begin{eqnarray}
f^{T}_{opt}\:(\rho_{nmems}^{AD}) = \frac{1}{2}+\nonumber{}\\
\frac{1}{9}\:\sqrt{\sin^{4}(\theta)p^{2}-2\:sin^{4}(\theta)p+ \sin^{4}(\theta)-2\:\sin^{2}(\theta)p^{2} + 4\:\sin^{2}(\theta)p-2\sin^{2}(\theta)+p^{2}-2\:p+1}+ \nonumber{}\\ \frac{1}{18}\sqrt{3\:sin^{4}(\theta)\:p^{2}+6\:\sin^{4}(\theta)\:p-6\sin^{2}(\theta)\:p^{2}-12\:sin^{2}(\theta)p+3\:p^{2}+6p}
\label{p_16}.
\end{eqnarray}
The following figure (i.e. Fig. $2$) shows how ADC affects optimal teleportation fidelity of the state (\ref{p_1}).
\begin{figure}[h!]
\begin{center}
\includegraphics[width=11cm]{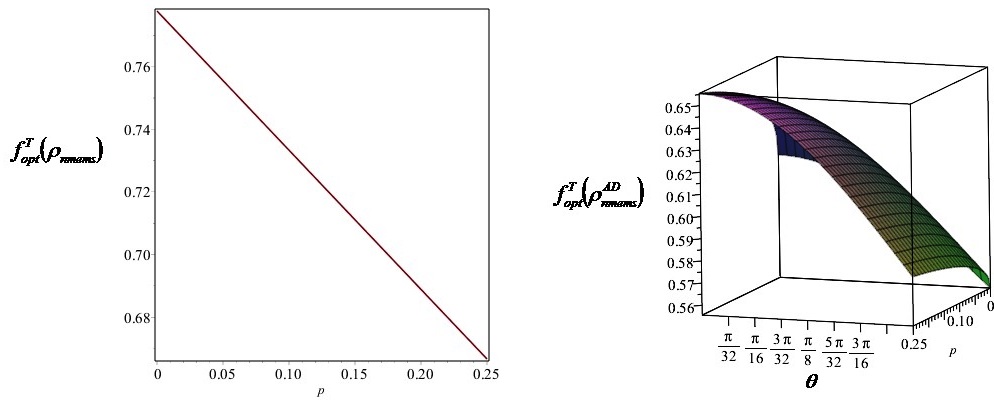}
\caption{(Coloronline) The first plot shows the original teleportation fidelity of the state $\rho_{nmems}$ where as the second plot gives the optimal teleportation fidelity of the state $\rho_{nmems}^{AD}$. In both the cases the state parameter $p$ is considered from $0$ to $0.25$.}
\end{center}
\label{f2}
\end{figure}
It is clear from the Fig. $2$ that the optimal teleportation fidelity of the state (\ref{p_1}) reaches its maximum at $0.777$ for $p=0$. After the ADC acts on this state, then the optimal teleportation fidelity of the transformed state (\ref{p_10}) reaches its maximum at $0.66$ for $p=0$ as well and  which is merely equal to the classical teleportation fidelity of $\frac{2}{3}$.\\\\
The quantum operation that describes the effect of dissipation to an environment at finite temperature is modelled as generalized amplitude damping channel. The generalized amplitude damping channel for a single qubit $\tau$ can be defined as \cite{nielsenbook}
\begin{eqnarray}
\tau = \sum_{i=0}^{3}\:E_{i}\:\tau\:E_{i}^{\dagger}
\label{p_17},
\end{eqnarray}
where,
\begin{eqnarray}
 E_{0}&=& \sqrt{\lambda}\left(%
\begin{array}{cc}
1 & 0  \\
0& \sqrt{1-\gamma} 
\end{array}%
\right),\:\:\:\:
 E_{1}= \sqrt{\lambda}\left(%
\begin{array}{cc}
0 & \sqrt{\gamma} \\
0& 0 
\end{array}%
\right)\nonumber{}\\
 E_{2}&=& \sqrt{1-\lambda}\left(%
\begin{array}{cc}
\sqrt{1-\gamma}  & 0  \\
0& 1 
\end{array}%
\right),\:\:\:\:
 E_{3}=  \sqrt{1-\lambda}\left(%
\begin{array}{cc}
0 & 0 \\
\sqrt{\gamma}& 0 
\end{array}%
\right)
\label{p_18}.
\end{eqnarray}
However we have skipped the discussion of generalized ADC here as ADC described in eqs. (\ref{p_7}) and (\ref{p_8}) and generalized ADC differ only in the location of the fixed point of flow \cite{nielsenbook} but their contributions on NMEMS are same i.e. both of these ADC's affect the fidelity of teleportation of the class of NMEMS.\\\\
It is known that, an arbitrary state $\tau_{ab}$ is classical if local measurements exist which do not perturb it. In other words one should have $\tau_{ab} = \tau_{ab}^{/}$ in this case. $`$Measurement Induced Disturbance' (in short MID) is an entropic cost of measurement in the basis of the states. This is defined as \cite{Modi2012}
\begin{eqnarray}
MID = S(\tau_{ab}^{/}) - S(\tau_{ab}).
\label{p_19}
\end{eqnarray}
$S(\tau_{ab})$ is the von-Neumann entropy of the state  and is given as \cite{nielsenbook}
\begin{eqnarray}
S(\tau) = −tr(\tau_{ab}\:\log\:\tau_{ab}) = - \sum_{i}\:k_{i}\:\log\:k_{i},
\label{p_20}
\end{eqnarray}
where $k_{i}$'s are the eigenvalues of the state $\tau_{ab}$ along with the condition that $0 \equiv 0 \log 0$. The MID for the constructed NMEMS (\ref{p_1}) is therefore found as 
\begin{eqnarray}
MID^{\rho_{nmems}} =  S(\rho_{nmems}^{AD}) - S(\rho_{nmems}) 
\label{p_20}
\end{eqnarray}
To better understand this we plot the expression of (\ref{p_20}) that is shown below.
\begin{figure}[h!]
\begin{center}
\includegraphics[width=11cm]{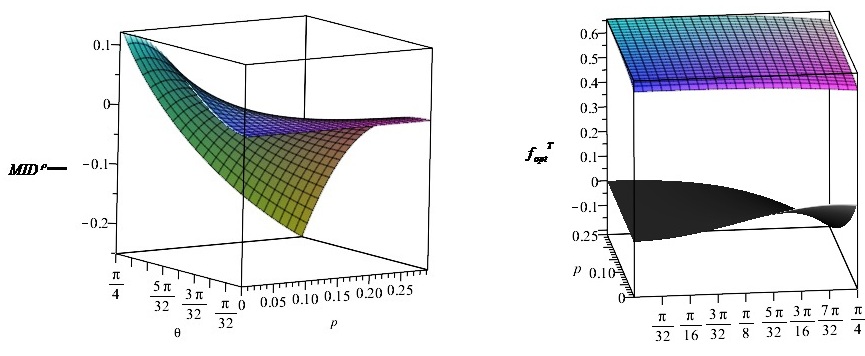}
\caption{(Coloronline) The figure shows the measurement induced disturbance on the state $\rho_{nmems}$ when amplitude damping channel acts on the state. The ordinate represents the MID and $0\:\leq p\:<\:0.292$, $0\:\leq \theta\:\leq \frac{\pi}{4}$. The second plot of this figure shows how measurement induced disturbance affects the optimal teleportation fidelity of the state $\rho_{nmems}^{AD}$.}
\end{center}
\label{f3}
\end{figure}
From the Fig. $3$ it is clear that, the measurement induced disturbance for the state (\ref{p_1}) reaches its maximum value when $p=0$ and $\theta = \frac{\pi}{4}$. To further analyse how the state's MID affected the optimal teleportation fidelity, we plot in Fig. $3$ the optimal teleportation fidelity of the state $\rho_{nmems}^{AD}$ as well as $MID^{\rho_{nmems}}$ together by varying the parameter in the range $[0,\:0.25)$ and the measurement angle $\theta$ in the range $[0,\:\frac{\pi}{4}]$, shown in the second plot of Fig. $3$.
\section{Discord and NMEMS:}
In quantum information theory, over the recent years, different methods for quantifying the quantum and classical parts of correlations emerged. Entanglement is one of the most prominent of all such measures of correlations. But there are instances where un-entangled states also exhibit non-classical behavior. Such ambiguities can be removed by developing some other appropriate entities which are capable of quantifying them. One of these measures is $`$Quantum Discord' \cite{Modi2012,ollivier2001}. Wang \textit{et. al} developed a way to quantify quantum discord for bipartite cases \cite{wang2011}. For an arbitrary bipartite state $\tau_{ab}$, an expression for quantum discord (or often simply known as discord, denoted as $QD$) is given by
\begin{eqnarray}
QD(\tau_{ab}) = \min(Q_{1},Q_{2})
\label{p_21},
\end{eqnarray}
where, 
\begin{eqnarray}
Q_{i} &=& H(\tau_{11}+\tau_{33}) + \sum_{i=1}^{4}\:\epsilon_{i}\log_{2}(\epsilon_{i}) + D_{j},\nonumber{}\\
D_{1} &=& H\lbrace\: \frac{1+\sqrt{[1-2\:(\tau_{33}+\tau_{44})]^{2}+4\:(\vert \tau_{14}\vert + \vert \tau_{23}\vert)^{2}}}{2}\:\rbrace,\nonumber{}\\
D_{2} &=& -\sum_{i}\:\tau_{ii}\:\log_{2}(\tau_{ii}) - H(\tau_{11}+\tau_{33}),\nonumber{}\\
H(x) &=& -x\:\log_{2}\:(x)-  (1-x)\:\log_{2}\:(1-x).\nonumber{}\\
\label{p_22}
\end{eqnarray}\\
Then using eq. (\ref{p_22}) we calculate the quantum discord for our constructed NMEMS $\rho_{nmems}$, denoted by $QD_{\rho_{nmems}}$
\begin{eqnarray}
QD_{\rho_{nmems}} = \nonumber{}\\
\min\:\lbrace\:-\frac{(p+2)x}{6\:\log\:2} + \frac{x\:\log(x)}{\log(2)} + \frac{y\:\log(y)}{\log(2)}-\frac{2\:z\log\:(z)}{\log(2)},\nonumber{}\\
\frac{(p-4)t}{6\:\log(2)} - \frac{(p+2)\log(x)}{6\:\log(2)}+\frac{p\:r}{\log(2)} + \frac{x\log(x)}{\log(2)} +\frac{y\:\log(y)}{\log(2)} - \frac{t_{1}\log(t_{1})}{\log(2)} - \frac{t_{2}\log(t_{2})}{\log(2)}\:\rbrace,
\label{p_23}
\end{eqnarray}
where,
\begin{eqnarray}
x=\frac{p+2}{6},\:\:\: y=\frac{2-2\:p}{3},\nonumber{}\\
z=\frac{1-p}{3},\:\:\: t= \frac{4-p}{6},\nonumber{}\\
r=\frac{p}{2},\:\:\: t_{1} = \frac{1}{2}+\frac{(1-p)\sqrt{5}}{6},\:\:\: t_{2} = \frac{1}{2}-\frac{(1-p)\sqrt{5}}{6}
\label{p_24}
\end{eqnarray}
Using eqs. (\ref{p_13}), (\ref{p_15}) and (\ref{p_23}) we plot below the concurrence, optimal teleportation fidelity and quantum discord for the state (\ref{p_1}) in the following figure.\\\\
\begin{figure}[h!]
\begin{center}
\includegraphics[width=11cm]{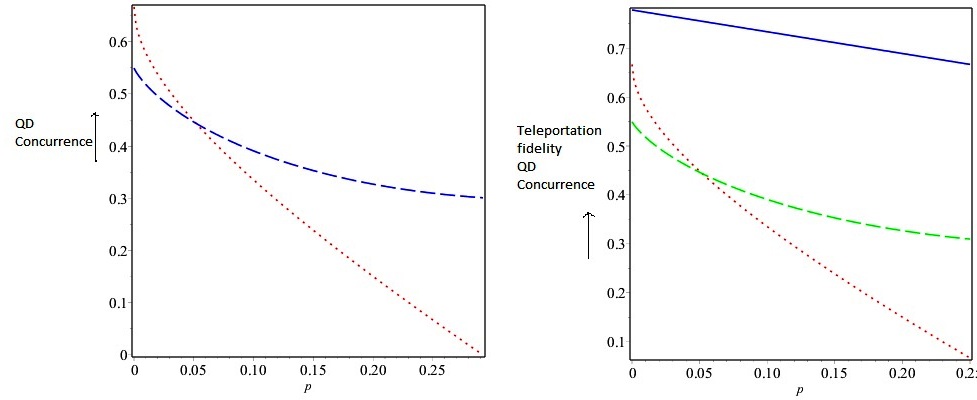}
\caption{(Coloronline) In the first plot quantum discord ($QD$) and the concurrence for $\rho_{nmems}$ are plotted. In the second plot optimal teleportation fidelity, quantum discord and concurrence for $\rho_{nmems}$ have been plotted. In the first plot the range of the state parameter $p$ lies between $0$ and $0.292$ while the range of $p$ is taken from $0$ to $0.25$ in the second plot.}
\end{center}
\label{f4}
\end{figure}\\
From the first plot of Fig. $4$ it is clear that both quantum discord and the concurrence are decreasing functions of the state parameter $p$. It is clear from the figure that for $p=0.06$, the quantum discord and the concurrence of the state (\ref{p_1}) are same. The decay of concurrence is faster than that of quantum discord. At $p=0$, i.e. where the state (\ref{p_1}) reduces to the form of Ishizaka - Hiroshima class of maximally entangled mixed states, the value of the concurrence of the state is higher than the quantum discord of the state. The concurrence of the state (\ref{p_1}) is approximately $0.66$ while the discord of the state is found to be $0.56$ for $p=0$. Moreover, for $0\:\leq p\: \leq\: 0.06$, the concurrence of the state (\ref{p_1}) is more than that of states' discord, while for $0.06\: < \: p \: < 0.292$, the concurrence of the state (\ref{p_1}) is less than quantum discord of the state.\newpage
\section{Conclusion:}
In this article we mainly focussed on three different aspects, the search for an appropriate witness operator for our constructed NMEMS, the NMEMS will be considered under noisy environment and the study of quantum discord of the NMEMS. We have found two witness operators, an entanglement witness shown in eq. (\ref{p_5}) and a teleportation witness in eq. (\ref{p_6}) for $\rho_{nmems}$. The effects of noisy environment on the NMEMS (\ref{p_1}) has also been studied. For this study we emphasized on amplitude damping channel. We have seen that the channel has significant influence on the NMEMS and the NMEMS (\ref{p_1}) is no longer useful as teleportation channel. The measurement induced disturbance of the state $\rho_{nmems}$, when amplitude damping channel acts on it has been calculated and it was also shown how MID is connected to optimal teleportation fidelity of the state (\ref{p_1}). The state (\ref{p_1}) has also been analyzed from the view point of quantum discord and a comparative study of the states' quantum discord and the states' concurrence as well as it's optimal teleportation fidelity have been made. We have seen that for a specified value of the parameter of the state  i.e. for specific $p$ the quantum discord and concurrence, both are having the same value. Moreover, for certain ranges of the state parameter, the concurrence is more than the quantum discord. In future, we can further extend our study of the NMEMS with respect to some other interesting noisy channels such as covariant channels, phase damping channels, de-polarizing channels. It is known that, in Cavity - QED analysis, one can generate non-maximally entangled mixed state, when two atoms pass through the cavity one after another. Basically by taking trace over the cavity field we get atom - atom NMEMS \cite{biplab2006,biplab2007}. The utilities of such generated NMEMS in quantum information processing tasks can be studied further. The noise channel effects on such generated states can also be studied. 

\section*{Acknowledgement:}
The authors S. Roy and B. Ghosh acknowledge their co-authors from their earlier paper of ref. \cite{roy2010}.

\end{document}